\documentstyle[preprint,aps]{revtex}
\tightenlines

\begin{document}

\title{Continuous Variable Quantum Cryptography}
\author{T.C.Ralph}
\address{Department of Physics, Faculty of
Science, \\ The Australian National University, \\ ACT 0200 Australia \\ 
Fax: +61 6 249 0741  Telephone: +61 6 249 4105 \\ E-mail: 
Timothy.Ralph@anu.edu.au}
\maketitle

%\begin{center}
%\scriptsize (26th April 1999)
%\end{center}

\begin{abstract}

We propose a quantum cryptographic scheme in which small phase and 
amplitude modulations of CW light beams 
carry the key information. The presence of EPR type correlations 
provides the quantum protection.

\end{abstract}

\vspace{10 mm}

Quantum cryptographic schemes use fundamental properties of quantum 
mechanics to ensure the protection of random number keys \cite{wie83,ben84}. 
In particular the act of measurement in quantum mechanics inevitably 
disturbs the system. Further more, for single quanta such as a photon, 
simultaneous measurements of non-commuting variables are forbidden. 
By randomly encoding the information between non-commuting 
observables of a 
stream of single photons any eavesdropper (Eve) is forced to guess 
which observable to measure for each photon. On average, half the time Eve 
will guess wrong, revealing her self through the back action of the 
measurement to the sender (Alice) and receiver (Bob). There are some 
disadvantages in working with single photons, particularly in 
free-space where scattered light levels can be high. Also it is of 
fundamental interest to quantum information research to investigate 
links between discrete variable, single photon phenomena and 
continuous variable, multi-photon effects. This motivates a 
consideration of quantum cryptography using multi-photon light modes. 
In particular we consider encoding key information as small signals 
carried on the amplitude and and phase quadrature amplitudes of the 
beam. These are the analogues of position and momentum for a light 
mode and hence are continuous, conjugate variables. 
Although simultaneous measurements of these non-commuting observables 
can be made in various ways, for example splitting the 
beam on a 50:50 beamsplitter and then making homodyne measurements on 
each beam, the information that can be obtained is 
strictly limited by the generalized uncertainty principle for 
simultaneous measurements \cite{yam86,aut88}. If an ideal measurement of one 
quadrature amplitude produces a result with a signal to noise of
\begin{equation}
(S/N)^{\pm}={{V_{s}^{\pm}}\over{V_{n}^{\pm}}}
\end{equation}
then a simultaneous measurement of both quadratures cannot 
give a signal to noise result in excess of
\begin{equation}
(S/N)_{sim}^{\pm}=({{\eta^{\pm}V_{s}^{\pm}}\over{\eta^{\pm}
V_{n}^{\pm}+\eta^{\mp}V_{m}^{\pm}}}) S/N^{\pm}
\label{sn}
\end{equation}
Here $V_{s}^{\pm}$ and $V_{n}^{\pm}$ are, respectively, 
the signal and noise power of 
the amplitude ($+$) or phase ($-$) 
quadrature at a particular rf frequency with respect to the 
optical carrier. The quantum noise which is inevitably added when dividing the 
mode is $V_{m}^{\pm}$. The splitting ratio is $\eta^{\pm}$ and 
$\eta^{+}=1-\eta^{-}$ (e.g a 50:50 beamsplitter has 
$\eta^{+}=\eta^{-}=0.5$).
The spectral powers are normalized to the quantum 
noise limit (QNL) such that a coherent beam has $V_{n}^{\pm}=1$. Normally 
the partition noise will also be at this limit ($V_{m}^{\pm}=1$). For 
a classical light field, i.e. where $V_{n}^{\pm}>>1$ the penalty will 
be negligible. However for a coherent beam a halving of the signal to 
noise for both quadratures is unavoidable when the splitting ratio is 
a half. The 
Hartley-Shannon law \cite{har49} applies to Gaussian, additive noise, 
communication channels such as we will consider here. It shows, in general, 
that if information of a fixed bandwidth is being 
sent down a communication channel at a 
rate corresponding to the channel capacity and the signal to noise is 
reduced, then errors will inevitably appear at the receiver. Thus, 
under such conditions, any attempt by an eavesdropper to make 
simultaneous measurements will introduce errors into the transmission. 
In the following we will 
first examine what level of security is guaranteed by this uncertainty 
principle if a coherent state mode 
is used. We will then show that the level of security can in 
principle be made as strong as for the single quanta case
by using a special type of 
two-mode squeezed state. The question of optimum protocols and 
eavesdropper strategies is complex and has been studied in detail for 
the single quanta case \cite{Fus96}. Here we only examine 
the most obvious strategies and do not attempt to prove equal security 
for all possible strategies.

Consider the set up depicted in Fig.1. A possible protocol is 
as follows. Alice generates two 
independent random strings of numbers and encodes one on the phase 
quadrature, and the other on the amplitude 
quadrature of a bright coherent beam. Bob uses homodyne detection to 
detect either the amplitude or phase quadrature of the beam when he 
receives it. He swaps randomly which quadrature he detects. On a 
public line Bob then tells Alice at which quadrature he was looking, at 
any particular time. They pick one quadrature to be the test and 
the other to be the key. For example, they may pick the amplitude 
quadrature as the test signal. They would then compare results for the times that 
Bob was looking at the amplitude quadrature. If Bob's results agreed 
with what Alice sent, to within some acceptable error rate, they would 
consider the transmission secure. They would then use the undisclosed 
phase quadrature signals, sent whilst Bob was observing the phase 
quadrature, as their key. By randomly swapping which quadrature is key 
and which is test throughout the data comparison an increased error 
rate on either quadrature will immediately be obvious.

To quantify our results we will consider the specific 
encoding scheme of binary pulse code modulation, in which the data is 
encoded as a train of 1 and 0 electrical pulses which are impressed on 
the optical beam at some RF frequency using electro-optic modulators. 
The amplitude and phase signals are imposed at the same frequency with 
equal power. Let us now consider 
what strategies Eve could adopt (see 
Fig.2). Eve 
could guess which quadrature Bob is going to measure and measure it 
herself (Fig.2(a)). 
She could then reproduce the digital signal of that 
quadrature and impress it on another coherent beam which she would 
send on to Bob. She would learn nothing about the other quadrature 
through her measurement and would have to guess her own random string 
of numbers to place on it. When Eve guesses the right quadrature to 
measure Bob and Alice will 
be none the wiser, however, on average 50\% of the time Eve will guess 
wrong. Then Bob will receive a random string from Eve unrelated to the 
one sent by Alice. These will agree only 50\% of the time. 
Thus Bob and Alice would see a 25\% bit error rate in the test 
transmission if Eve was using this strategy. This is analogous to the 
result for single quanta schemes in which this type of strategy 
is the only available. 

However for bright beams it is possible to make simultaneous 
measurements of the quadratures, with the caveat that there will be 
some loss of information. So a second strategy that Eve could follow 
would be to split the beam in half, measure both quadratures and 
impose the information obtained on the respective quadratures of 
another coherent beam which she sends to Bob (Fig.2(b)). 
How well will this 
strategy work? Suppose Alice wishes to send the data to Bob with a 
bit error rate (BER) of about 1\%. For bandwidth limited transmission 
of binary pulse code modulation \cite{yar97}
\begin{equation}
BER={{1}\over{2}}erfc{{1}\over{2}}\sqrt{{{1}\over{2}}S/N}
\label{ber}
\end{equation}
Thus Alice must impose her data with a S/N of about 
13dB. For simultaneous measurements of a coherent state the signal to 
noise obtained is halved (see Eq.\ref{sn}). As a result, using 
Eq.\ref{ber}, we find the information Eve intercepts and subsequently 
passes on to Bob will only have a BER of 6\%. This is clearly a superior 
strategy and would be less easily detected. Further more Eve could 
adopt a third strategy of only intercepting a small amount of the 
beam and doing simultaneous detection on it (Fig.2(c)). For 
example, by intercepting 16\% of the beam, Eve could gain information 
about both quadratures with a BER of 25\% whilst Bob and Alice would 
observe only a small increase of their BER to 1.7\%. In other words Eve 
could obtain about the same amount of information about the key that she 
could obtain using the ``guessing'' strategy, whilst being very 
difficult to detect, especially in the presence of losses.

The preceding discussion has shown that a cryptographic scheme based 
on coherent light provides much less security than single quanta 
schemes \cite{note1}. We now consider whether squeezed light can offer improved 
security. For example amplitude 
squeezed beams have the property $V_{n}^{+}<1<V_{n}^{-}$. 
Because the amplitude quadrature 
is sub-QNL greater degradation of S/N than the coherent case 
occurs in simultaneous 
measurements of amplitude signals (see Eq.\ref{sn}). Unfortunately the 
phase quadrature must be super-QNL, thus there is less degradation of 
S/N for phase signals. As a result the total security is in fact less 
than for a coherent beam. However in the following we will show that by
using two squeezed light beams, security comparable to that achieved 
with single quanta can be obtained. 

The set-up is shown in Fig.3. Once again Alice encodes her number 
strings digitally, but now she impresses them on the amplitude 
quadratures of two, phase locked, 
amplitude squeezed beams, $a$ and $b$, one on each. A $\pi/2$ phase 
shift is imposed on beam $b$ and then they are mixed on a 
50:50 beamsplitter. The resulting output modes, $c$ and $d$, are given by
\begin{eqnarray}
c & = & \sqrt{{{1}\over{2}}}(a+i b)\nonumber\\
d & = & \sqrt{{{1}\over{2}}}(a-i b)
\end{eqnarray}
These beams are now in an entangled state which will exhibit Einstein, 
Podolsky, Rosen (EPR) type correlations \cite{ein35,ral98}. Local 
oscillator beams (LO's) of the same power as, and 
with their polarizations rotated to be orthogonal to 
$c$ and $d$ are then mixed with the beams on polarizing 
beamsplitters. A rapidly varying random time delay is imposed on one 
of the beams. Both mixed beams 
are then transmitted to Bob who uses polarizing beamsplitters to 
extract the local oscillator from each beam. Bob {\it cannot} remix 
the signal beams ($c$ and $d$) 
to separate $a$ and $b$ because the random time delay introduced
between the beams has destroyed their coherence at the signal frequency. However, 
because each beam has a corresponding local oscillator which has 
suffered the same time delays, 
Bob {\it can} make individual, phase sensitive measurements on each 
of the beams and extract either the information on $a$ or the 
information on $b$ by amplifying the local oscillators and using 
balanced homodyne detection. Note that the noise of the LO's is 
increased by amplification 
but balanced homodyne detection is insensitive to LO noise.
He randomly chooses to either: 
(i) measure the 
amplitude quadratures of each beam and add them together, in which case
he obtains the power spectrum
\begin{eqnarray}
V^{+} & = & <|(\tilde c^{\dagger}+\tilde c)+(\tilde d^{\dagger}
+\tilde d)|^{2}>\nonumber\\
 & = & V_{s,a}+V_{n,a}^{+}
\end{eqnarray}
where the tildes indicate Fourier transforms. Thus he obtains the data string 
impressed on beam $a$, $V_{s,a}$, 
imposed on the sub-QNL noise floor of beam $a$, $V_{n,a}^{+}$, or; 
(ii) measure the phase quadratures of each 
beam and subtract them, in which case he obtains the power spectrum
\begin{eqnarray}
V^{-} & = & <|(\tilde c^{\dagger}-\tilde c)-(\tilde 
d^{\dagger}-\tilde d)|^{2}>\nonumber\\
 & = & V_{s,b}+V_{n,b}^{+}
\end{eqnarray}
i.e. he obtains the data string impressed on beam $b$, $V_{s,b}$, 
imposed on the sub-QNL noise floor of beam $b$, $V_{n,b}^{+}$. Thus 
the signals lie on conjugate quadratures but 
{\it both} have sub-QNL noise floors. This is the hallmark of the EPR 
correlation \cite{ou92}.

Consider now eavesdropper strategies. Firstly, like Bob, Eve cannot remix $c$ 
and $d$ optically to obtain $a$ and $b$ due to the randomly varying phase 
shift ($\phi(t)$) introduced by the time delay. For small phase shifts 
beam $c$ becomes $c'=(a+ib)(1+i \phi)$. Mixing $c'$ and $d$ on a 
beamsplitter will produce outputs with amplitude power spectra
\begin{eqnarray}
V_{c'+d} & = & V_{s,a}+V_{n,a}^{+}+\alpha^{2} V_{\phi} \nonumber\\
V_{c'-d} & = & V_{s,b}+V_{n,b}^{+}+\alpha^{2} V_{\phi}
\end{eqnarray}
where $\alpha^{2}$ is proportional to the intensity of beams $a$ and 
$b$ and $V_{\phi}$ is the power spectrum of the phase fluctuations. If 
$\phi(t)$ has a white power 
spectrum over frequencies from well below to well above the signal 
frequency the signals will be obscured. It is not possible to directly 
control the phase shifts with 
out similarly suppressing the signals. However the phase shifts are 
also present on the LO co-propagating with $c'$. Mixing the two 
LO's will produce an output with amplitude power spectra
\begin{equation}
V_{+LO}=1+E^{2}V_{\phi}
\end{equation}
where $E^{2}$ is proportional to the intensity of the LO's and the 
``one'' is from the quantum noise of the LO's. It is possible to use 
this output to control the phase noise on the mixed signal beams 
giving (ideally) the amplitude power spectra
\begin{eqnarray}
V_{c'+d}^{C} & = & V_{s,a}+V_{n,a}^{+}+{{\alpha^{2}}\over{E^{2}}} \nonumber\\
V_{c'-d}^{C} & = & V_{s,b}+V_{n,b}^{+}+{{\alpha^{2}}\over{E^{2}}}
\label{control}
\end{eqnarray}
where the remaining penalty arises from the quantum noise of the LO's.
If $E^{2}>>\alpha^{2}$ (as is normally the case for a LO)
then this penalty can be made negligible, thus retrieving the 
signals. This is why it is essential that the LO's have the same power as the 
signal beams at the point where the phase fluctuations are imposed. 
This makes the ratio of the correlated phase noise to the 
independent quantum noise the same for the LO and the signal beam. 
This cannot be changed by Eve. 
With $E^{2}=\alpha^{2}$ the penalty is at the quantum limit. 
As we shall see in a moment this is sufficient to reveal Eve.
  
Eve can still adopt the guessing strategy by detecting a particular 
quadrature of both beams and then using a similar apparatus to Alice's 
to re-send the beams. As before she will only guess right half the 
time thus introducing a BER of 25\%. Suppose instead she tries the 
second strategy of simultaneous detection of both quadratures on each 
beam. She will obtain the 
following power spectra for the summed amplitude quadratures and the 
differenced phase quadratures.
\begin{eqnarray}
V^{+} & = & {{1}\over{2}}(V_{s,a}+V_{n,a}^{+}+1)\nonumber\\
V^{-} & = & {{1}\over{2}}(V_{s,b}+V_{n,b}^{+}+1)
\label{twin}
\end{eqnarray}
The signal to noise is reduced as predicted by Eq.\ref{sn} 
but where the noise power for both quadrature measurements is 
sub-QNL \cite{note3}. This leads to improved security. For example with 10 dB 
squeezing ($V_{n,a}=V_{n,b}=0.1$) the signal to noise in a 
simultaneous measurement will be reduced by a factor of .09. 
As a result,assuming initial S/N of 13dB and using 
Eq.\ref{ber}, we find the information Eve intercepts and subsequently 
passes on to Bob will now have a BER of about 24\%. In other words, the 
security against an eavesdropper using simultaneous measurements 
is now on a par with the guessing strategy. The third strategy is also 
now of no use to Eve as small samples of the fields carry virtually no 
information. For example, with 10 dB squeezing, intercepting 16\% of 
the field will give Eve virtually no information (a BER of 49.5\%) 
whilst already producing a 5\% BER in Bob and Alice's shared 
information.

In any realistic situation losses will be present. Losses tend in 
general to reduce security in quantum cryptographic schemes 
\cite{bar96}. The 
problem for our system is that losses force Alice to increase her 
initial S/N in order to pass the information to Bob with a low BER. 
Eve can take advantage of this by setting up very close to Alice. 
Never-the-less reasonable security can be maintained with sufficiently 
high levels of squeezing. For example with 10 dB squeezing and 10\% 
loss,
strategy two will result in a 15\% BER in the shared information. Also 
Eve must intercept 29\% of the light to obtain a 25\% BER 
using the third strategy which will cause a 20\% BER in Alice and 
Bob's information. With 6 dB squeezing and 20\% loss the second strategy 
penalty is reduced to a BER of 7.5\%, similar to that of the coherent 
state scheme. However, for the third strategy, 
Eve must still intercept 29\% of the light to 
obtain a BER of 25\% and this will cause an 11\% BER in Alice and 
Bob's shared information, much larger than for the coherent case. 
Although these results demonstrate some tolerance to loss for our 
continuous variable system it should be 
noted that single quanta schemes can tolerate much higher losses 
\cite{butt98} 
making them more practical from this point of view.   

In summary we have examined the quantum cryptographic security of two 
continuous variable schemes, one based on coherent light, the other 
based on 2-mode squeezed light. Whilst the coherent light scheme is 
clearly inferior to single quanta schemes, the squeezed light scheme 
offers, in principle, equivalent security. The quantum security is 
provided by the generalized uncertainty relation. It is also 
essential that the coherence between the two squeezed modes is 
destroyed. 
More generally this system is an example of a 
new quantum information technology based on continuous variable, 
multi-photon manipulations. Such technologies may herald a new 
approach to quantum information.

\begin{figure}
\caption{Schematic of coherent light cryptographic set-up. AM is an 
 amplitude modulator whilst PM is a phase modulator.}
\end{figure}

\begin{figure}
\caption{Schematics of three eavesdropper strategies. Only the (a) 
 is available in single quanta schemes. }
\end{figure}

\begin{figure}
\caption{Schematic of squeezed light cryptographic set-up. Sqza and 
 sqzb are phase locked squeezed light sources. Rna and Rnb are 
 independent random number sources. Bs and pbs are non-polarizing and 
 polarizing beamsplitters respectively. Half-wave plates to rotate the 
 polarizations are indicated by $\lambda/2$ and optical amplification 
 by $A$. The $\pi/2$ phase shift 
 is also indicated. HD stands for homodyne detection system.}
\end{figure}


\begin{thebibliography}{99}
\bibitem{note1} Another strategy Eve could use is to do homodyne 
detection at a quadrature angle half-way between phase and amplitude. 
This fails because the signals become mixed. Thus Eve can tell when 
both signals are 0 or both are 1 but she cannot tell the 
difference between 1,0 and 0,1. This again leads to a 25\% BER.
\bibitem{note3} The signal to noise properties of Eq.~\ref{control} 
are the same as those of Eq.~\ref{twin}.
\end{thebibliography}
\end{document}